\documentclass[12pt]{article}
\usepackage{amstext}
\usepackage{bm}
\usepackage{amssymb,amsmath,mathrsfs,setspace}
\usepackage{graphicx,psfrag,epsf}
\usepackage{array}
\usepackage{natbib}
\usepackage{epsfig}
\usepackage{amssymb}
\usepackage{url}
\usepackage{listings}
\usepackage{verbatim}
\usepackage{comment}
\usepackage{color}
\usepackage{graphicx,psfrag,subfigure,amsmath,amssymb,appendix,multirow,float,threeparttable,abstract}

\usepackage[top=1in, bottom=1in, left=1in, right=1in]{geometry}
\fontsize{12}{0}
\numberwithin{equation}{section}
\date{}

\begin{document}

	\title{\bf FastLORS: Joint Modeling for eQTL Mapping in R \thanks{Address for correspondence: X. Jessie Jeng, Department of Statistics, North Carolina State University, SAS Hall, 2311 Stinson Dr., Raleigh, NC 27695-8203, USA. E-mail: xjjeng@ncsu.edu. 
		}\hspace{.2cm}}
	\author{Jacob Rhyne$^1$, Eric Chi$^1$, Jung-Ying Tzeng$^{1}$,  and X. Jessie Jeng$^1$ \\ 
		1. Department of Statistics, North Carolina State University\\
		}
	\maketitle

\begin{abstract}
\textbf{Motivation:} Yang et al. (2013) introduced LORS, a method that jointly models the expression of genes, SNPs, and hidden factors for eQTL mapping. LORS solves a convex optimization problem and has guaranteed convergence. However, it can be computationally expensive for large datasets. 
In this paper we introduce Fast-LORS which uses the proximal gradient method to solve the LORS problem with significantly reduced computational burden. \\
\textbf{Results:} We apply Fast-LORS and LORS to data from the third phase of the International HapMap Project and obtain  comparable results. Nevertheless, Fast-LORS shows substantial computational improvement compared to LORS. \\
\textbf{Availability:} All functions are available in the FastLORS R package, available at \url{https://github.com/jdrhyne2/FastLORS} \\
\textbf{Contact:} {xjjeng@ncsu.edu}
\end{abstract}

\maketitle

\section{Introduction}

Expression quantitative trait loci (eQTLs) are genomic regions that carry DNA sequence variants that influence gene expression.  In eQTL mapping, associations are identified between genetic variants, usually single nucleotide polymorphisms (SNPs), and the expression of genes \citep{review2015}.  Early methods of eQTL mapping involved marginal tests for each SNP and gene pair \citep{Matrix_eQTL}.  More recent approaches utilize joint modeling to account for the effects of multiple SNPs on the expression of a gene; examples include \cite{StatMethods}, \cite{PANAMA}, \cite{LMMEH}, \cite{yang2013LORS}, etc.  \cite{yang2013LORS} introduced a LOw-Rank representation to account for confounding factors and make use of Sparse regression for eQTL mapping (LORS).  LORS has many attractive features, including (1) it handles the modeling of hidden factors, which has been shown to increase both \textit{cis} and \textit{trans}-eQTL detection \citep{PANAMA}, (2) it does not assume independence amongst genes, (3) the optimization problem for sparse regression is convex, and (4) with proper choice of tuning parameters, the algorithm is guaranteed to converge \citep{yang2013LORS}.

Consider a dataset with $n$ individuals with expression levels of $q$ genes and genotypes of $p$ SNPs.  Let $\textbf{Y}$ be an $n\times q$ matrix of gene expression levels, $\textbf{X}$ be an $n\times p$ matrix of SNPs, $\textbf{B}$ be an $p\times q$ matrix of coefficients, $\textbf{L}$ be an $n\times q$ matrix of hidden factors, $\textbf{$\mu$}$ be a $1\times q$ vector of intercepts, and $\textbf{e}$ be an $n\times q$ matrix of Gaussian random error.  The LORS model is defined as
\vspace{-0.1in}
\begin{equation}
\label{LORS_Model}
\textbf{Y} = \textbf{1$\mu$} + \textbf{XB} + \textbf{L} + \textbf{e}.
\end{equation}

It is assumed that there are only a few hidden factors that influence gene expression and that gene expression levels are affected by only a small fraction of SNPs \citep{yang2013LORS}.  In other words, it is assumed that $\textbf{L}$ is low-rank and $\textbf{B}$ is sparse.  The LORS optimization problem is

\vspace{-0.2in}
\begin{equation}
\label{LORS_opt}
min_{\textbf{B},\textbf{$\mu$},\textbf{L}} \frac{1}{2} \lVert \textbf{Y} - \textbf{XB} - \textbf{1} \textbf{$\mu$} - \textbf{L} \rVert^2_F + \rho \lVert \textbf{B} \rVert_1 + \lambda \lVert \textbf{L} \rVert_*.
\end{equation}

\noindent where the term $\rho\lVert \textbf{B} \rVert_1$ enforces the sparsity of $\textbf{B}$ and $\lambda \lVert \textbf{L} \rVert_*$ with nuclear norm enforces the low-rank constraint of $\textbf{L}$.  \cite{yang2013LORS} solves the optimization problem in (\ref{LORS_opt}) through an alternating procedure in which $\textbf{B}$ and $\textbf{$\mu$}$ are held fixed and $\textbf{L}$ is solved for (the $\textbf{L}$-step) and then $\textbf{L}$ is held fixed while $\textbf{B}$ and $\textbf{$\mu$}$ are solved for (the $\textbf{B},\textbf{$\mu$}$-step).  For a given $\textbf{Y}$, $\textbf{X}$, $\rho$, and $\lambda$, the LORS algorithm updates $\textbf{B}$, \textbf{$\mu$}, and $\textbf{L}$ by

\begin{enumerate}
\item $\textbf{L}^+ \leftarrow SVD(\textbf{Y} - \textbf{XB} - \textbf{1$\mu$}, \lambda)$;
\item $\textbf{B}_j^+, \textbf{$\mu$}_j^+ \leftarrow min_{B_j,\mu_j} \frac{1}{2} \lVert \textbf{Y}_j - \textbf{XB}_j - \textbf{1$\mu$}_j - \textbf{L}_j \rVert^2_F + \rho \lVert \textbf{B}_j \rVert_1,$
\end{enumerate}

\noindent where $SVD(.,\lambda)$ refers to the soft-thresholded singular value decomposition and $j=1,...,q$.  While this algorithm will converge for a properly chosen $\rho$ and $\lambda$, it is computationally expensive for datasets with a large number of genes or SNPs.  In this paper, we introduce a new algorithm named Fast-LORS to  solve the optimization problem in (\ref{LORS_opt}) by the proximal gradient method.

Fast-LORS has the same alternating structure as LORS, but uses different approaches to update $\textbf{B}$, \textbf{$\mu$}, and $\textbf{L}$ by 

\begin{enumerate}
\item $\textbf{L}^+ \leftarrow prox_{t_L, h = \lambda \lVert \rVert_*}(\textbf{L} - t_L \triangledown g(\textbf{L}; \textbf{B},\textbf{$\mu$})) $;
\item $\textbf{B}^+ \leftarrow prox_{t_B, h = \rho \lVert \rVert_1}(\textbf{B} - t_{B} \triangledown g(\textbf{B};\textbf{$\mu$},\textbf{L})) $;
\item $\textbf{$\mu$}^+ \leftarrow \textbf{$\mu$} - t_\mu \triangledown g(\textbf{$\mu$}; \textbf{B}, \textbf{L})$,
\end{enumerate}

\noindent where $prox_{th}(.)$ is the proximal mapping function \citep{prox_lit}, $g(\textbf{B}, \textbf{$\mu$}, \textbf{L}) = \frac{1}{2} \lVert \textbf{Y} - \textbf{XB} - \textbf{1$\mu$} - \textbf{L} \rVert $, 
and $t_L,t_B$, and $t_\mu$ are step-sizes.  Fast-LORS shows a sharp decrease in the objective function values over the initial iterations, and the objective function values can rapidly approach the optimal value at convergence.  If the objective function values are not strictly decreasing, the updating steps of Fast-LORS will revert to the updating steps of LORS.  

Fast-LORS eliminates the need to solve $q$ independent lasso problems in Step 2 of the LORS algorithm, which has complexity $O(n^2pq)$ \citep{yang2013LORS}.  In Step 2 of our algorithm, the dominant step is the multiplication of the matrices in the calculation of the gradient, which has complexity $O(npq)$.  Thus, our algorithm reduces the computational complexity of the B-step by a factor of $n$.

Our R package gives the user the option to solve the optimization problem in (\ref{LORS_opt}) using either Fast-LORS or LORS algorithm.  Current availability of LORS algorithm is exclusive to MATLAB. We provide the LORS in R, which is translated from the MATLAB package available at \url{http://zhaocenter.org/software/}.

\section{Example}

The functions required to run the Fast-LORS and LORS algorithms are provided in the FastLORS R package.  The major function in the package is Run\_LORS.  The required inputs are the user defined matrices of $\textbf{Y}$ and $\textbf{X}$.  Two optional arguments of interest are for joint modeling and pre-screening. 

Suppose the user is handling a dataset with  a large number of SNPs. He/she may elect to use pre-screening to reduce the number of SNPs before joint modeling is performed.  We include two options for pre-screening: the "LORS-Screening" introduced in \cite{yang2013LORS} and the HC-Screening recently developed in \cite{2018arXiv180402737R}. Comparing to LORS-Screening, HC-Screening shows promise in retaining SNPs for both cis- and trans- eQTLs detection \citep{2018arXiv180402737R}. 
Run\_LORS can apply Fast-LORS with HC-Screening by

\begin{lstlisting}[language=R]
> Run_LORS(Y, X, method = "Fast-LORS", screening = "HC-Screening")
\end{lstlisting}

If the user wishes to run the original LORS algorithm for joint modeling and LORS-Screening for pre-screening, this can be accomplished by setting the method argument to "LORS" and the the screening argument to "LORS-Screening".


An additional argument of note is the cross\_valid argument, the default of which is TRUE.  \cite{yang2013LORS} do not perform cross-validation in their parameter tuning procedure.  In practice, we have observed that the sparsity of $\textbf{B}$ varies with different training and testing sets, thus by default Run\_LORS performs two-fold cross validation to tune the parameters.  The example to apply the original LORS algorithm without cross-validation and use LORS-Screening to reduce the number of SNPs before joint modeling is 

\begin{lstlisting}[language=R, breaklines = true]
> Run_LORS(Y, X, method = "LORS", screening = "LORS-Screening", cross_valid = FALSE)
\end{lstlisting}.

If the data is small enough so that pre-screening is not needed the screening argument can simply be set to "None".  The Run\_LORS function returns (1) the estimated $\textbf{B}$, $\textbf{L}$, and $\textbf{$\mu$}$ (2) the objective function values (3) the relative change in objective function values (4) the number of iterates (5) the time required for screening, parameter tuning, and joint modeling, respectively.  The nonzero entries of the estimated $\textbf{B}$ are the eQTLs identified by Fast-LORS or LORS.

\section{Results}

We acquired SNP and gene expression data from the Asian populations from the third phase of the International HapMap Project (HapMap3).  The expression data is available at \url{http://www.ebi.ac.uk/arrayexpress/experiments/E-MTAB-264/} and the SNP data is available at \url{ftp://ftp.ncbi.nlm.nih.gov/hapmap/genotypes/hapmap3_r3/plink_format/}.  We used HC-Screening to reduce the number of SNPs, applied two-fold cross validation to tune $\rho$ and $\lambda$, and used Fast-LORS and LORS to perform joint modeling on the reduced dataset. Though the default maximum number of iterations is 100 for LORS, we increased the maximum iterations of both methods to ensure that both algorithms converged.  A summary of the time required for each modeling method and the number of eQTLs detected is presented in Table 1.

\begin{table}[ht]
\begin{center}
\caption{Time Required for Joint Modeling and eQTLs Detected}
\begin{tabular}{  l | l | l }
  \hline			
  Method  & Time (min) & eQTLs Detected \\
  \hline
  Fast-LORS  & 1.65 & 178 \\
  LORS  & 79.49 & 181 \\
  \hline  
\end{tabular}
\end{center}
\end{table}

As seen in Table 1, Fast-LORS takes dramatically less time to run than LORS and identifies a comparable number of eQTLs.  Additionally, of the 181 eQTLs identified by the LORS algorithm, 176 are also identified by Fast-LORS.  Thus, Fast-LORS produces very similar results to LORS and greatly reduces the computational time required for joint modeling.  

We also compared Fast-LORS with HC-Screening to the original LORS with LORS-Screening.  The total time required--including pre-screening, parameter tuning, and joint modeling--is approximately 7 hours and 27 minutes for Fast-LORS with HC-Screening and approximately 19 hours and 38 minutes for the original LORS with LORS-Screening.  Therefore, by using Fast-LORS in conjunction with HC-Screening,  the computational time is reduced by almost twelve hours compared to the original LORS screening and modeling methods.

\section{Conclusion}

The FastLORS R package gives users the ability to perform eQTL mapping for a gene expression matrix and a matrix of SNPs through joint modeling.  Fast-LORS improves upon the speed of joint modeling without much sacrifice in eQTL detection and HC-Screening largely reduces the computation time by pre-screening while retaining important eQTLs. The use of the two methods together in the FastLORS R package allows a user to perform efficient joint modeling for eQTL mapping.

\section*{Funding}

Dr. Jeng was supported by National Human Genome Research Institute of the National Institute of Health under grant R03HG008642. Dr. Tzeng was partially supported by National Institutes of Health grant P01 CA142538.

%
%

\end{document}